\documentclass[12pt]{article}

\usepackage[english]{babel}
\usepackage{amssymb}
\usepackage{graphicx}
\usepackage{cite}
\usepackage{subfigure}
\def\beq{\begin{equation}}
\def\eeq{\end{equation}}
\def\bea{\begin{eqnarray}}
\def\eea{\end{eqnarray}}
\def\nn{\nonumber}
\def\sN{\mathbf{\bar N}}
\def\sL{\mathbf{L}}
\def\sH{\mathbf{H_u}}
\def\sHd{\mathbf{H_d}}
\def\nn{\nonumber}
\def\supot{\mathcal{W}}
\def\nL{n_L^{(L)}}
\def\nR{n_L^{(R)}}
\def\sneu{\bar{\tilde{\nu}}}

\begin{document}


\author{Steven Abel and V\'eronique Pag\'e \\
    {\em Centre for Particle Theory,} \\
{\em University of Durham, Durham, DH1 3LE, UK}}

\title{Affleck-Dine (Pseudo)-Dirac Neutrinogenesis}

\maketitle

\abstract
\noindent We consider the Affleck-Dine mechanism for leptogenesis 
in the minimal MSSM with Dirac or Pseudo-Dirac neutrinos.  
The rolling of scalars along $D$-flat 
directions generates a left-right asymmetry in
the sneutrino sector, only the left part of which is transferred to a
baryon asymmetry via sphaleron transitions.  In the pure Dirac case
the baryon asymmetry of
the Universe is thus mirrored by an equal and opposite asymmetry in the
leptons. The mechanism is also found to work when the neutrinos are 
pseudo-Dirac. No additional field needs to be added to the MSSM 
other than the right-handed neutrino.

\section{Introduction}
It was noticed some time ago that leptogenesis, far from
requiring lepton-number violating Majorana masses
as in the original scenario \cite{leptogen}, can in fact be
implemented in the SM with purely Dirac neutrinos \cite{dicketal}.
This mechanism, called \emph{neutrinogenesis}, 
relies on the fact that
$(B+L)$-violating transitions leave the right-handed sector unaffected
\cite{shaposh}; as long as left-right equilibrating processes are
small enough to be out of equilibrium, 
it is possible to 'hide' a right-handed
lepton asymmetry from the sphaleron transitions. The idea of hiding 
lepton number in inert species has a long history 
\cite{davidsoncampbellolive} but
works particularly effectively for the neutrinos; indeed
it was shown in
\cite{dicketal} that Dirac neutrinos easily satisfy this condition.
The temporary left-handed lepton asymmetry can thus be processed before
the electroweak phase transition into today's observed baryon
asymmetry. It is only well after the phase transition that the neutrinos' 
Yukawa couplings come into equilibrium, by which time  
the sphalerons are quenched and the baryon asymmetry is locked in.
The usefulness of this idea lies in the fact that Dirac 
neutrinos of the right size can arise in models where GUT scale  
degrees of freedom are integrated out because Yukawa couplings in 
such models can be naturally suppressed by factors of $M_W/M_{GUT}$. 
This possibility has received increased interest recently 
in the context of supergravity and effective models of string theories 
\cite{arkanihamed,borzumatinomura,casasetal,kitano,arnowittetal,abelsakis}. 
The fact that baryogenesis is also possible then leaves open the 
intriguing possibility that $B-L$ is conserved in Nature or that 
neutrinos are in fact pseudo-Dirac rather than Majorana.

However the toy model used in \cite{dicketal} used
an additional heavy Higgs-like doublet, because the scenario 
worked by 'drift and decay' as in original leptogenesis. 
In the present paper we point out that the Affleck-Dine (AD)
mechanism \cite{admech} allows an extremely efficient 
implemententation of neutrinogenesis in just the MSSM with 
Dirac neutrinos.  A
$(\tilde{\nu}_L-\tilde{\nu}_R)$ current can be produced through the
rolling of scalars along their $D$-flat directions; although
lepton-number is conserved, only left-handed lepton number 
can be converted to a baryon number through sphalerons,
and the right-handed component is hidden by the smallness of the Yukawa
coupling as before.  The $B+L$ number of the Universe is thus mirrored by an
equal and opposite right-handed lepton number, until the right-handed
(s)neutrino oscillations decay long after the electroweak phase transition. 
We should mention that 
AD neutrinogenesis was proposed in ref.\cite{balaji}. However 
in that work the AD field was considered to be an additional
scalar field that was either Higgs-like, with $SU(2)$ number, 
or a singlet appearing in higher order non-renormalizable interactions. 
The implementation here using only the
$D$-flat directions of the MSSM itself
can be thought of as the minimal realisation of AD neutrinogenesis 
in the context of supersymmetry. 
Apart from testing whether AD can work in a $B-L$ preserving 
MSSM this minimal scenario is naturally very predictive,
because for example the CP violating potential 
arises from soft-supersymmetry breaking 
trilinear terms, the coupling of the AD field to the 
neutrinos is given by the neutrino mass, and so on.

We first discuss the evolution equations for the
left-right ($L-R$) asymmetry as a result of the Dirac mass term added
to the MSSM superpotential.  We will then evolve numerically the $L-R$
asymmetry; using the equilibrium relations between $B$ and $L$, this
asymmetry will be converted to a baryon asymmetry. 
Finally we discuss the pseudo-Dirac and 'weak see-saw' cases. If 
neutrinogenesis is to work a bound on the Yukawa couplings results 
from the requirement that the oscillations of the AD field decay 
after the electroweak phase transition.  
The nett result is a constraint on the size of the additional Majorana 
mass which turns out to be 
\[
M_R \lesssim 0.6 \left( \frac{0.05 \mathrm{eV}}{m_\nu} \right) 
\mathrm{MeV}.
\]
This bound is the strongest, being slightly more severe than the 
requirement that the baryon number generated be large enough. 
In short this means that the 
AD neutrinogenesis is able to operate for 
all 'reasonable' Dirac and pseudo-Dirac scenarios in the neutrino 
mass sector. It even works for mild see-saw cases although the 
latter are probably excluded by nucleosynthesis. 

\section{The Superpotential}
Let us first introduce the right-handed neutrino superfield, 
$\mathbf{\bar N}$, and add a Dirac mass term for the neutrinos in the 
superpotential:
\beq \label{superpotential}
\mathbf{\mathcal{V_S}} \supset  \lambda \sL^i \epsilon_{ij} \sH^j \sN ~,
\eeq
where $\mathbf{L}$ is the left-handed lepton doublet and $\mathbf{H_u}$ is the up-type Higgs.  
The gauge invariants $\sL \sH$ and $\sN$ will be the important $D$-flat directions 
\cite{drandallt1}; let us parameterise them as
\bea
 L &=& \frac{1}{\sqrt{2}} \left( \begin{array}{c} \phi \\ 0 \end{array} \right) ~, \nn\\
 H_u &=& \frac{1}{\sqrt{2}} \left( \begin{array}{c} 0 \\ \phi \end{array} \right) ~, \nn\\
 \bar N  &=& \bar{\widetilde{\nu}}
\eea
where non-bold letters stand for the scalar part of the superfields.
The (SUSY-conserving) scalar potential arising from the added Dirac
mass term is
\bea
\mathcal{V}_F &=& \left| \frac{\partial \supot}{\partial L^a}\right|^2 +  \left| \frac{\partial \supot}{\partial H^b}\right|^2 + \left| \frac{\partial \supot}{\partial \bar{\nu}}\right|^2  \\
&=& \frac{\left|\lambda\right|^2}{4}\left|\phi^2\right|^2 + \left|\lambda\right|^2 \left| \sneu \phi \right|^2 ~.
\eea
These tiny $F$-term contributions lift the 
 $\sL \sH$ and $\sN$ flat directions very slightly. 
In usual AD the directions considered would be 
both $D$ {\em and} $F$-flat at the renormalizable level, and the 
flat direction would be lifted only by non-renormalizable 
(i.e. higher dimension)
operators. Here the directions would be $F$-flat as well 
but for the neutrino mass. It is only because of the smallness of the 
latter that we can hope to send the field out to large enough 
VEVs to generate asymmetries. Note that conversely when we go on later to 
consider pseudo-Dirac neutrinos, then the scenario begins to 
run into difficulty if there is any significant 
see-saw effect at work in the neutrino mass matrices: a significant 
see-saw would imply larger Dirac Yukawa couplings and lift these 
directions more. 

The AD mechanism of course requires additional CP violation, and here 
it comes from the soft-breaking sector;
\beq
\mathcal{V}_{SB} = m_{\phi}^2 \left|\phi\right|^2 + m_{\sneu}^2 \left|\sneu \right|^2 + (\lambda a \phi^2 \sneu + h.c.)~.
\eeq
As was discussed in \cite{drandallt1,drandallt2}, soft-breaking terms also
get a contribution from 
the non-zero Hubble constant in the early Universe and this is crucial as
it drives the fields out to large values during inflation. 
We parameterise these as
\beq \label{hubblemass}
\mathcal{V}_{H} = -c_{\phi}H^2 \left|\phi\right|^2 -c_{\nu} \left|\bar{\nu} \right|^2 + (\lambda c_H H \phi^2 \sneu + h.c.)~.
\eeq
The overall potential for the scalar fields is thus
\bea \label{potential}
\mathcal{V} &=& \mathcal{V}_F +\mathcal{V}_{SB} + \mathcal{V}_H \nn\\
 &=& (m_{\phi}^2-c_{\phi}H^2) \left|\phi\right|^2 + (m_{\sneu}^2 -c_{\nu} H^2) \left|\sneu \right|^2 + (\lambda (a+c_H H) \phi^2 \sneu + \mathrm{h.c.}) \nn\\
 &+& \frac{\left|\lambda\right|^2}{4}\left|\phi^2\right|^2 + \left|\lambda\right|^2 \left| \sneu \phi \right|^2~.
\eea
For the flat directions to develop large expectations values during inflation, 
we need at least one of the fields to have a negative effective mass squared; here we will consider
\beq
(m_{\phi}^2-c_{\phi}H^2)<0 ~.
\eeq
Thus the Hubble induced terms in eq.(\ref{hubblemass}) push the fields
far from the origin. They are also important in introducing a
time-dependence into the potential which guarantees that the
AD mechanism will be in operation. To see this note that
without these terms the potential always has a minimum where 
sign$(\lambda a \phi^2 \bar{\tilde{\nu}})=-1$ as this is the only trilinear
term. Without an initial kick the fields will simply roll down this
valley and no nett lepton currents of any kind will be generated.  The
Hubble induced terms mean that the minimum is now at sign$(\lambda
(a+c_H H) \phi^2 \bar{\tilde{\nu}})=-1$. Thus even if $c_H$ is real the
phases of the fields have to become time dependent to track the
instantaneous minimum: in effect the Hubble constant should kick the
AD field for us wherever it starts out.

\section{The dynamics of the asymmetry}
To see these effects we proceed to examine how the asymmetry develops. 
First write the lepton number $n_L$ as a sum of its right-handed and 
left-handed parts:
\beq \label{lnbr}
n_L=n_L^{(L)}+n_L^{(R)}
\eeq
with $n_L^{(L)}$ and $n_L^{(R)}$ being in terms of our scalar fields
\bea \label{nlnr}
\nL &=& \frac{i}{2}\left( \dot{\phi}^*\phi -\phi^*\dot{\phi} \right) \nn\\
\nR &=& -i \left(\dot{\sneu}^*\sneu-\sneu^*\dot{\sneu} \right) ~.
\eea
The evolution equation for $\phi$ is:
\beq \label{srphi}
\ddot{\phi} + 3 H \dot{\phi} + \frac{\partial \mathcal{V}}{\partial \phi^*}=0
\eeq
and analogously for $\sneu$.  Now using eq.(\ref{nlnr}) in eq.(\ref{srphi}) 
and its conjugate, we find
\beq
\dot{n}_L^{(L)} + 3 H \nL =\mathrm{Im}
\left( \frac{\partial \mathcal{V}}{\partial \phi} \phi \right),
\eeq
and again analogously for $\sneu$.
From eq.(\ref{potential}) we see that the only imaginary terms are the $a$-terms and hence
\bea
\dot{n}_L^{(L)}+3H\nL& =& 2 \mathrm{Im}\left( \lambda a \phi^2 \sneu \right) \nn\\
\dot{n}_L^{(R)}+3H\nR &=& -2 \mathrm{Im}\left( \lambda a \phi^2 \sneu \right) ~.
\eea
We can see that lepton number eq.(\ref{lnbr}) is conserved,
\beq
\dot{n}_L+3Hn_L = \frac{d}{dt}\left( \nL+\nR \right)+3H\left(\nL+\nR \right) = 0,
\eeq
but that the left-right asymmetry, $\nL-\nR \equiv n_{LR}$, has a non-trivial evolution:
\beq \label{lrasy}
\dot{n}_{LR}+3Hn_{LR} = 4 \mathrm{Im} \left( \lambda a \phi^2 \sneu \right)~.
\eeq

The hope then is that this time dependence and CP asymmetry in the 
potential will generate a nett $n_{LR}$. If it does so 
this will feed through to the baryons via sphalerons. Before 
continuing we briefly establish the relation between $n_L^{(R)}$ 
and the baryon number 
(before taking account of the effect of sphalerons, 
$n_{L}^{(R)}$ is initially half the 
value of $n_{LR}$ generated during the evolution of
eq.(\ref{lrasy})): the equilibrium ratio between 
lepton and baryon number under
rapid sphaleron transitions was 
calculated in ref.\cite{harveyturner,dreinerross} for an SM like structure. 
In the present case we have an out-of-equilibrium
right-handed Dirac neutrino in the analysis which simply holds a
nett $B-L$ and remains completely inert; therefore we can set 
$(B-\hat{L})=n_L^{(R)}$ where $\hat{L}$ is the sum of all the 
leptons in equilibrium (i.e. excluding the right handed neutrino).
In the MSSM we have an additional charged Higgs which changes the 
result from the SM; repeating the chemical potential analysis and assigning a chemical potential 
$\mu_{B-\hat{L}}$ we find that above the electroweak phase transition 
for $m$ Higgs fields 
\begin{eqnarray}
Y & = & 16 \mu_{B-\hat{L}} + (20+2m) \mu_Y\nonumber \\
B-\hat{L} & = & 26 \mu_{B-\hat{L}} + 16 \mu_Y
\end{eqnarray}
while below it 
\begin{eqnarray}
Q & = & 16 \mu_{B-\hat{L}} + (44+4m) \mu_Y\nonumber \\
B-\hat{L} & = & 26 \mu_{B-\hat{L}} + 16 \mu_Q.
\end{eqnarray}
Imposing $(B-\hat{L})=n_L^{(R)}$ this then translates into
the following ratios:
\bea
\label{harveyturner}
\begin{array}{cccccccc}
B &= & L & = & \frac{4(6+m)}{66+13m}n_{L}^{(R)} & = & \frac{8}{23}n_{L}^{(R)}  &  \qquad T > T_{ew}, \\ 
B &=& L &  = &\frac{4(9+m)}{111+13m}n_{L}^{(R)} & = & \frac{44}{137}n_{L}^{(R)} &  \qquad T < T_{ew},\\
\end{array} 
\eea
where $L$ is the total lepton number including the right handed neutrinos and $m=2$ in the 
MSSM. (The only effect of the charged Higgs of the MSSM is to change the denominator
of the last expression to $111+13m $ rather than $98+13m$.) Note that 
in this pure Dirac case $B-L$ is conserved and the 
final baryon number is approximately the LSP density, given by 
$n_{DM}=\frac{23}{8}n_B $, reminiscent of ideas pursued in 
a number of works \cite{tytgat,dmfarrar,dmbarr1,dmbarr2,dmkaplan,dmkuzmin,dmkitano1,dmkitano2,dmthomas,dmkusenko,dmhooper}. This 
suggests the right handed sneutrino as the preferred LSP candidate 
since the LSP mass would then have to be of order 1GeV:
\beq
m_{DM} = \frac{8}{23}\frac{\Omega_{DM}}{\Omega_{b}}m_b\, .  
\eeq

Returning now to the dynamical evolution first note that 
the forcing term for $n_{LR}$ evolution is time dependent
and this is true generically when $H\gg m_{3/2}$ 
even if the fields are sitting in the 
minimum because the minimum of the potential depends on $H(t)$. 
Indeed we have to minimise
\bea
\mathcal{V}  &=& (m_{\phi}^2-c_{\phi}H^2) \left|\phi\right|^2 + (m_{\sneu}^2 -c_{\nu} H^2) \left|\sneu \right|^2 -2 \left|\lambda (a+c_H H)\right| \left|\phi\right|^2 \left| \sneu \right|  \nn\\
 &+& \left|\lambda\right|^2\left|\phi^2\right|^2 + \left|\lambda\right|^2 \left| \sneu \phi \right|^2~.
\eea
Taking the coefficient of $\left| \sneu \right|$ positive, and 
for $\left| c_{\nu} \right|H^2 \gg m_{\nu}^2$ and  $\left| c_{\phi} \right|H^2 \gg m_{\phi}^2$, the minimum of the potential is given by
\beq \label{minimum}
\left|\phi\right|_{\mathrm{min}}(t) \simeq \sqrt{\frac{c_{\phi}}{2}}\frac{H(t)}{\lambda} \nn 
\eeq
\begin{displaymath} 
\left|\sneu\right|_{\mathrm{min}}(t) \simeq \left\{ \begin{array}{ll}
\frac{-c_{\phi}}{2c_{\nu}-c_{\phi}} \frac{\left| a \right|}{\left| \lambda \right|}, & c_H H \ll \left| a\right|, \\ 
\frac{-c_{\phi}}{2c_{\nu}-c_{\phi}} \frac{\left| H(t) \right|}{\left| \lambda \right|},  & c_H H \gg \left|a\right|.
\end{array}  \right. 
\end{displaymath} 
One obvious difference between the present case and the more usual 
one in \cite{drandallt1} is that here the initial field values have a lower 
bound of order $\phi_{\mathrm{min}} \simeq a/\lambda \sim 10^{14}$ GeV even if  
$H$ itself is much smaller than this value.

Let us now sketch the evolution from early after inflation to the
electro-weak phase transition; we will confirm the picture 
with a numerical solution of the equations of motion as we 
go along. The numerically evolving $n_{LR}$ is shown in fig.(1).

The mechanism requires that initially
the fields are drawn far along the flat directions during inflation.
This in turn requires the coefficient of $\left| \phi \right|^2$ in
eq.(\ref{potential}) to be negative as we have seen above.  
Inflation is then followed by an
era of inflaton oscillation, during which the Universe is
matter-dominated ($H \sim 2/3t$). At this early stage the 
time dependence of the Hubble constant is still important in the 
evolution of the fields. 
Indeed both $\phi$ and $\sneu$ are following their evolution equations
and, since $H \gg m_{\phi}, m_{\sneu}$, their motion is dominated by 
the falling value of the $H^2 $ Hubble induced mass-squared terms.

The behaviour of the fields in this phase is a major difference  
between the scenario we are considering here and the original AD 
scenario. The flat direction here is lifted by renormalizable terms; 
it is easy to see from the equations of motion that
the distance of the fields from the instantaneous 
minimum drops as $t^{-1}$; but eq.(\ref{minimum}) tells us that the 
 minimum itself drops as $t^{-1}$ as well, so that during this matter dominated 
phase the fields are relatively undamped. We expect to see a long 
transient period and as our numerical analysis shows, this is 
indeed the case. To get a crude understanding of the behaviour 
in this phase, we can estimate the maximum amplitudes of the 
fields by assuming that the energy is constant in a co-moving volume: 
$R^3 H^2 \phi_{\mathrm{max}}^2=const$. This gives $\phi_{\mathrm{max}}=const$ which in turn gives 
$n_{LR} = const$. The detailed behaviour of $n_{LR}$ is still rather 
complicated at this point; the fields are not yet executing 
regular cycles, and the current $n_{LR}$ is rapidly flipping 
sign as the fields change their sense of rotation around the origin. 
This behaviour is in agreement with the arguments of 
refs.\cite{drandallt1} where it was noted that only 
with nonrenormalizable terms of dimension 4 or higher do the 
AD fields follow the instantaneous mininum closely. 

It is during matter-domination that the important 
$H\sim m_{3/2}$ mark is reached. 
Below this point the Hubble induced terms in the effective potential become 
irrelevant to the evolution which is now dominated by the mass terms,
and the behaviour changes markedly. 
The time dependence of the fields becomes more amenable to analytic
approximaton now since the equations of motion are nearly linear; 
indeed we may immediately use constancy of energy in a comoving volume 
argument to infer that $R^3 m^2 \phi_{\mathrm{max}}^2 = const$
where $m$ is the mass of whichever field we are considering. 
If $H=b/t$ this then suggests
\bea
\phi_{\mathrm{max}}\sim t^{-\frac{3b}{2}},
\eea
which we indeed confirm numerically. In matter domination $b=2/3 $ 
so that $n_{LR}$ drops as $t^{-1}$. This can clearly be seen in fig.(1). 
Slightly later, but 
before reheating, the fields begin to exhibit
``canonical'' AD behaviour where they {\em do} execute
regular cycles. Here we may approximate the real and imaginary components of fields 
as $t^k \sin(m t)$. Linearizing the equations of motion we find
\bea 
\frac{k(k-1) \sin(m_\phi t)}{t^2}+
\frac{2 k m_\phi \cos(m_\phi t)}{t}  
+\frac{3 b m_\phi \cos(m_\phi t)}{t} = 0,
\eea
and then neglecting $1/t^2 $ terms we find $k=-3b/2$ agreeing with the 
above.

The fields stay in this phase until reheating when the
Universe becomes radiation dominated. We assume that reheating happens 
for $T_R = 10^9$ GeV when the Hubble constant is $H\sim T_R^2/M_{Pl} \sim 1$ GeV. 
We now have $H=1/2t$, which gives $k=-3/4$,  
in accord with the late time behaviour of the Bessel function solutions of 
ref.\cite{admech}. The current then drops as $n_{LR}\sim t^{-3/2}$ as regular matter
until the electroweak phase transition.
This behaviour is again confirmed
by the numerical treatment: in particular fig.(1) shows
the initial transients, the switch to AD behaviour, the 
$t^{-1}$ and the $t^{-3/2}$ behaviour. 

\begin{center}
\begin{figure}
\begin{center}
\begin{minipage}{0.9\textwidth} 
\includegraphics[width=0.9\textwidth]{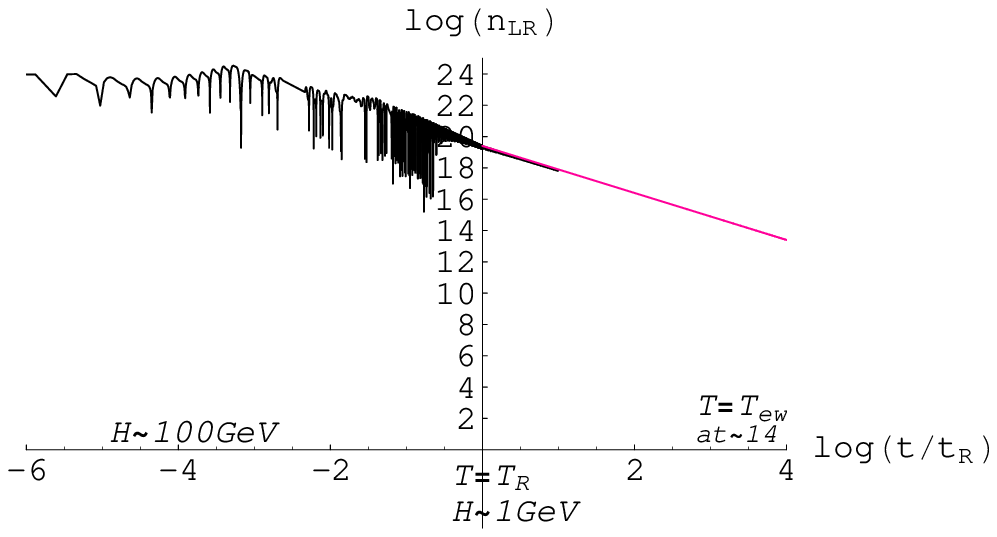}
\end{minipage}
\end{center}
\begin{center}
\begin{tabular}{c c}
\includegraphics[width=0.4\textwidth]{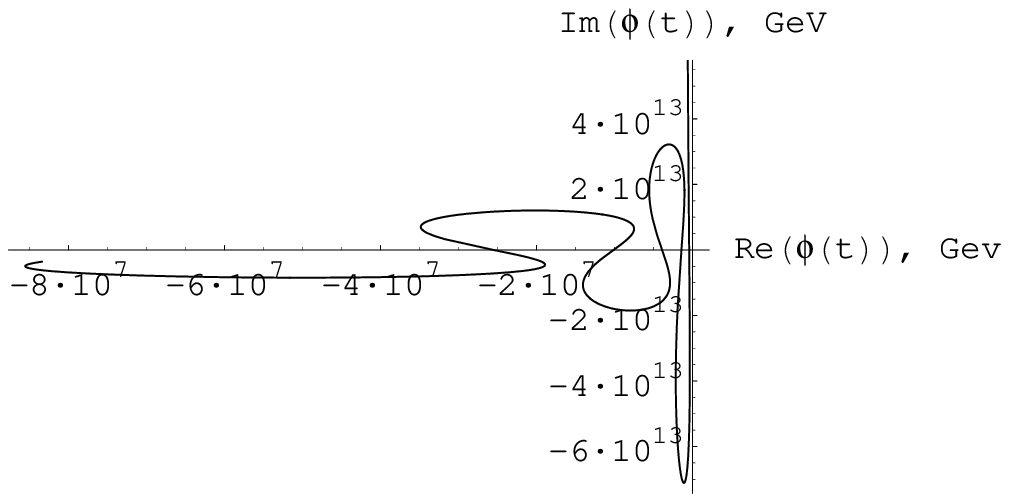}
 &
\includegraphics[width=0.4\textwidth]{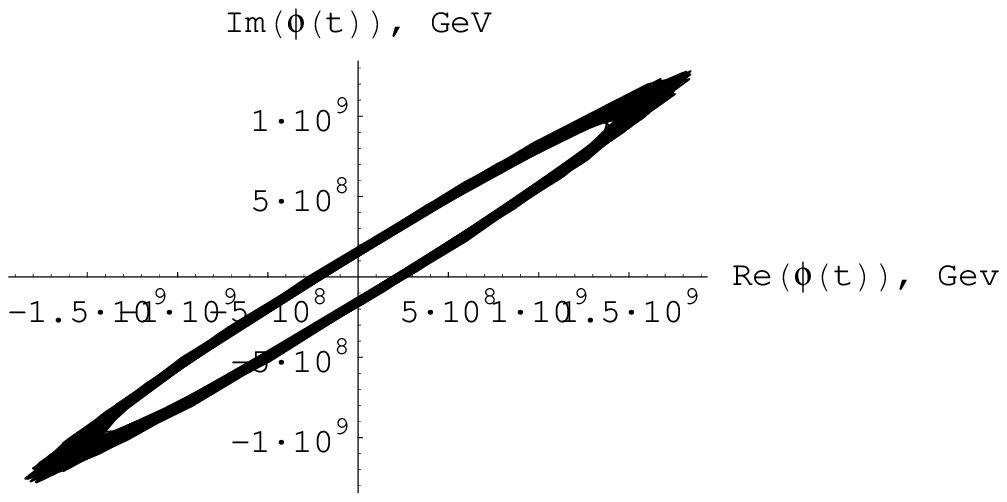}
\end{tabular}
\end{center}
\caption{Time evolution of the generated LR asymmetry.   Parameters and initial
conditions are as follows: $m_\phi=600$ GeV, $m_{\sneu}=500$ GeV,
$a=e^{0.6i}100$ GeV, $c_\phi=1$, $c_{\sneu}=0.8$, $c_H=0$,
$\lambda=10^{-12}$, $\phi(t_{in})=\imath \left|\phi\right|_{\mathrm{min}}(t_{in})$, 
$\sneu(t_{in})=\left|\sneu\right|_{\mathrm{min}}(t_{in})$,
$\dot{\phi}=\dot{\sneu}=0$, where the minima are given by the expressions 
in the text.  The added line is matter evolution during radiation 
domination, $t^{-3/2}$.  The behaviour of the $\phi$ field is also shown for early (shortly before $H \sim 100$GeV) and late (post-reheating) times.}
\label{fig:baryonnum}
\end{figure}
\end{center}

It is in this final phase 
as the fields are rolling down to the minimum, that they capture a 
left-right asymmetry that is constant relative to the entropy.  A positive
$n_{LR}$ means for instance that there is instantaneously more
left-handed sneutrinos and right-handed anti-sneutrinos than
left-handed anti-sneutrinos and right-handed sneutrinos, respectively:
the left-handed sneutrinos are quickly turned into left-handed
neutrinos through gaugino interactions. 
This can either go by decay with $\Gamma \sim g_2^2 m_{\sneu} $ 
or at high temperatures by scattering whose rate is
\beq
\Gamma \sim \frac{g_2^4}{m_{\tilde{W},\tilde{B}}^4}T^5\, 
\eeq
where the masses are understood to be thermal ones. 
All of the contributions are of the same order during the period we are 
considering when $T\sim M_W$ and so sneutrino$\leftrightarrow$neutrino 
conversion is in equilibrium. The sphaleron transitions transfer 
the left-handed neutrino asymmetry
into a baryon asymmetry as described above.  
Above the electroweak phase transition 
this happens on a timescale of order TeV$^{-1}$ which is essentially
instantaneous; after the electroweak transition
the sphalerons are switched off and the non-zero baryon number 
is frozen in \cite{shaposh,cohenkn,troddenrio}. Throughout, the 
right-handed (s)neutrinos remain inert 
until their oscillations decay through the coupling
to the neutralino. We should therefore ensure that this happens 
at a temperature much lower than the electroweak transition 
temperature $T_{ew}$: the life-time of the sneutrino decay is 
\bea
\tau_{\tilde{\nu}_R} 
\simeq \frac{4\pi}{\lambda^2} \frac{1}{m_{\tilde{\nu}}}B_{\mathrm{Higgs}}
\eea
where $B_{\mathrm{Higgs}}<1$ is the fraction of the neutralino that is
made of the up Higgs. To be conservative we take $B=1$, and find
that $\tau_{\tilde{\nu}_R}\gtrsim H^{-1} $ for all $T\gtrsim \left(\frac{\lambda}{10^{-12}}\right) 
100$ MeV.  
Note for later use that the mechanism stops working when 
\beq
\label{bound1}
\lambda \gtrsim 10^{-9}
\eeq
 because the oscillations are damped before 
the electroweak phase transition takes place.

The discsussion above of course assumes 
that the LSP is the usual neutralino. In the possibility we've mentionned above 
that the right-handed sneutrino itself be 
the LSP, the decay time for the 
sneutrino oscillations is even later. 

\section{The baryon asymmetry}

Having established the dynamical behaviour of the fields in some detail, let us 
now turn to the requirements to successfully generate a baryon number.
First we have seen that the oscillations in $n_{LR}$ remain constant 
for an initial transient period when the inflaton oscillations are 
scaling like matter. They begin to scale like matter as well 
once the Hubble constant reaches $\sim m_{3/2}$. 
Therefore in order to evaluate the eventual $n_{LR}$ asymmetry 
it is most useful to consider 
the relative densities when $H\sim m_{3/2}$;  
the density in coherent inflaton oscillations is 
$\rho_I \sim H^2 M_{P}^2 \sim m_{3/2}^2 M_P^2$. 
At this stage the 
field VEVs are of order $\phi,\sneu \sim |a/\lambda|$ as in eq.({\ref{minimum}),
so that the energy density in their oscillations is of order 
$\rho_{\phi,\sneu}\sim m_{3/2}^2 |a/\lambda|^2 $. Since the latter also 
behaves  like regular matter, we can use it to keep track of $n_{LR}$
until the time of reheating:
\bea 
\frac{\rho_{\phi,\sneu}}{\rho_I} \sim \frac{  |a/\lambda|^2}{M_P^2}\, .
\eea
From reheating onwards it is the ratio with entropy that remains constant.
Since $\rho_{\phi,\sneu} = m_{\phi,\sneu} n_{\phi,\sneu}$ that is given by
\bea
\label{final}
\frac{n_{\phi,\sneu}}{s} & \approx & \frac{  |a/\lambda|^2}{M_P^2} \frac{T_R}{m_\phi} \nn \\
 & = & 10^{-9} 
\left| \frac{  a}{100 \mathrm{GeV}} \right|^2
\left| \frac{  10^{-12}}{\lambda} \right|^2
\left( \frac{  T_R}{1 \mathrm{TeV}} \right)
\left( \frac{100 \mathrm{GeV}}{m_\phi} \right)
\, .
\eea

\section{The pseudo-Dirac and mild see-saw cases}
It is interesting to implement this scenario in the
more general case where Majorana mass terms
are included in the superpotential (\ref{superpotential}):
\beq \label{superpotential2}
\mathcal{W} \supset 
\sN \lambda \sL^a \epsilon_{ab} \sH^b + M_R \sN \sN + \frac{M_L}
{\langle h^0_u \rangle^2} (\sL \sH)^2 .
\eeq
Such additional terms can arise in the same manner as the 
Dirac terms in the supergravity scenarios considered in ref.\cite{arkanihamed,borzumatinomura,casasetal,kitano,arnowittetal,abelsakis}; 
essentially the pure Dirac models require symmetries to prevent Majorana masses for the right-handed neutrinos 
that can be relaxed to allow non-renormalizable operators such as $\sH \sHd \sN \sN/M_{GUT} $.
For example such an operator could lead to left-handed 
Majorana masses $M_L \sim 3\times 10^{-5}-7\times 10^{-4}$ eV 
in the models considered in ref.\cite{abelsakis}. In order
to present as general a discussion as possible we will consider $M_{L,R}$ to be arbitrary 
parameters and consider the question of {\em when AD neutrinogenesis can work}. 
We will
mainly focus on $M_R$ since $M_L \gtrsim \lambda v$ would give 6 active 
neutrinos which is certainly ruled out by nucleosynthesis; other than this 
we will consider $M_{L,R}$ to be free parameters. Throughout the following discussion we shall assume that mass-squared differences 
given by measured neutrino oscillations are indicative of the actual masses. 

The most immediate concern is how the new terms could affect the classical dynamics. 
Assuming for the moment that $M_L=0$, the superpotential leads to the 
scalar potential (\ref{potential}):
\bea \label{potential2}
\mathcal{V}  &=& (m_{\phi}^2-c_{\phi}H^2) \left|\phi\right|^2 + (m_{\sneu}^2 -c_N H^2+
4M_R^2) \left|\sneu \right|^2 \nn\\
&+& (\lambda (a+c_H H) \phi^2 \sneu +\lambda M_R \phi^2 \sneu^* + \mathrm{h.c.}) \nn\\
 &+& \frac{\left|\lambda\right|^2}{4}\left|\phi^2\right|^2 + \left|\lambda\right|^2 \left| \sneu \phi \right|^2~.
\eea
Clearly a new trilinear term has appeared which, following
eq.(\ref{srphi}), could affect
the dynamics of the fields, and thus of the asymmetry if $M_R \gtrsim a$. However 
as we shall see such large values are not relevant for the AD scenario here. 

Next, the lepton-number violating interactions introduced by the Majorana mass have
to be constrained such that they do not erase the asymmetry.  The rate 
of these interactions when $M_R\neq 0$ is given by an exchange with $\nu_R $ which for $T\gg M_R$ is   
\beq
\Gamma_{LV} \simeq \frac{\lambda^4 M_R^2}{T} ~,
\eeq
Demanding that this rate be smaller than $H$ for the duration of neutrinogenesis 
imposes
\beq \label{neumass0}
\lambda^4 M_R^2 \lesssim \frac{T_{ew}^3}{M_{Pl}}  
\eeq
which taking $T_{ew} = 100 $GeV gives 
\beq \label{neumass}
\lambda^2 M_R \lesssim 3 \times 10^{-7} \mathrm{GeV.}
\eeq
As we shall see the resulting bounds are not very constraining. 
For $M_L$ the lepton number violating exchanges are 
now suppressed only by the gauge couplings, $g^4_2$, rather than $\lambda^4$; however 
this still gives an uninteresting bound $M_L \lesssim 0.3 \mathrm{MeV.}$

A different and more constraining bound comes from the Dirac Yukawas themselves; 
again considering $M_L=0$, 
the light neutrino mass, given by
\beq \label{neumasseqn}
m_{\nu} = \frac{\sqrt{M_R^2+(2\lambda v)^2}-M_R}{2} ~,
\eeq
can require a $\lambda$ substantially larger than $10^{-12}$ 
due to see-saw 
effects once the Majorana contributions become dominant; any bound on 
$\lambda $ coupled with the estimated neutrino mass puts an 
indirect bound on $M_R$. One such bound comes from the fact that if 
$\lambda \gtrsim 10^{-8}$
then left- and right-handed neutrino can equilibrate
above the electroweak phase transition, destroying any left-right
asymmetry \cite{dicketal}. However as we saw 
a stricter bound,
\beq
\label{bound2}
\lambda \lesssim 10^{-9},
\eeq
comes from the requirement that the oscillations remain undamped until 
after the electroweak phase transition. In addition we of course 
require that the produced baryon number is sufficient; from eq.(\ref{final})
we see that this is least constraining for maximum reheat temperature. 
Assuming that the gravitino reheat bound is satisfied $T_R\lesssim 10^{9}$GeV
and that $n_B/s\approx 10^{-10}$ we find 
the bound is (coincidentally) almost the same as in eq.(\ref{bound2}),
$\lambda \lesssim 3\times 10^{-9}$.
The added Majorana mass $M_R$ must therefore respect both
the constraints in (\ref{neumass}) and (\ref{bound2}). It is the 
latter which is most constraining; it gives 
\beq
\label{seesaw}
M_R \lesssim 0.6 \left( \frac{0.05 \mathrm{eV}}{m_\nu} \right) 
\mathrm{MeV}.
\eeq
This bound then prevents the see-saw mechanism from operating 
in  all its glory, but a weak see-saw effect may still be present
in the neutrinos while still allowing AD neutrinogenesis
to work with these flat directions. In addition note that the 
bound means that the evolution is virtually unnaffected by the presence of 
of the Majorana terms since $M_R$ is indeed much smaller than both $H$ 
and the trilinear term $a$ driving the dynamics.

In this context, following eq.(\ref{final}), there are two options for explaining
why $n_B/s\approx 10^{-10}$. The first is the possibility 
that $M_R \sim 1$MeV and that there is a weak see-saw mechanism in operation.
This seems rather unnatural since it introduces an additional 
mass-scale that itself requires explanation.  Moreover, it has been argued that
a $1$MeV sterile neutrino with such a (relatively) large mixing angle 
($\sin^{2}2\theta=2\times 10^{-7}$) would cause large amounts of 
sterile neutrino dark matter to be produced \cite{abazajian}.
This scenario would then be forbidden by overclosure 
(and possibly other cosmological constraints - see \cite{abazajian}).  
In such a case, the remaining possibility is that the reheat 
temperature was of order $1$TeV\footnote{It has been noted that such a 
low reheating temperature could arise in supersymmetry \cite{allah}, if one of 
the $D$ {\em and} $F$-flat directions acquires a VEV of order the Planck scale. 
(The flat directions required for the mechanism here do not get large enough 
VEVs to affect the reheat temperature by more than an order of magnitude).}.

\section{Conclusion}
In this paper we presented a minimal version of
neutrinogenesis with Dirac sneutrinos in the MSSM,
and showed that it can generate the observed baryon
asymmetry of the Universe.  No new fields need to
be added to the MSSM apart from right-handed neutrinos.  
The mechanism works by first generating a $\nu_L-\nu_R$ asymmetry
using the AD mechanism, with $D$-flat directions
involving sneutrinos and Higgses playing the roll of
the Affleck-Dine fields.  The flat directions are appropriately
lifted during inflation by the inclusion of finite-energy density
SUSY-breaking terms which drives the VEVs to large values.  
As long as left-right equilibration  
is out of equilibrium before the electroweak phase transition
(resulting in a bound on the Dirac neutrino Yukawas), 
the nett left-handed lepton number can 
drive sphaleron transitions and ultimately 
create the observed baryon asymmetry. 
We also showed that the conditions on
the smallness of the Yukawa couplings 
still allows the mechanism to be implemented for
pseudo-Dirac neutrinos, and can in fact support a 
weak see-saw mechanism.

\section{Acknowledgments}
The work of V.P. is funded by the Fonds Nature et Technologies du Qu\'ebec.
We thank Sacha Davidson for discussions at the beginning 
of this project. We are also grateful to K.~Hamaguchi, R.~Kitano, 
C.~Lunardini and M.~Ratz for comparison of results and for discussions, and 
to A.Mazumdar for useful comments.

\bibliographystyle{unsrt}
\bibliography{addiraclep7}

\begin{thebibliography}{10}

\bibitem{leptogen}
M.~Fukugita and T.~Yanagida.
\newblock {\em Phys. Lett.}, B174:45, 1986.

\bibitem{dicketal}
K.~Dick, M.~Lindner, M.~Ratz, and D.~Wright.
\newblock {\em Phys. Rev. Lett.}, 84:4039--4042, 2000.

\bibitem{shaposh}
V.A. Kuzmin, V.A. Rubakov, and M.E. Shaposhnikov.
\newblock {\em Phys.Lett.B}, 155(36), 1985.

\bibitem{davidsoncampbellolive}
B.~A. Campbell, S.~Davidson, and K.~A. Olive.
\newblock 
\newblock {\em Nucl. Phys.}, B399:111--136, 1993.

\bibitem{arkanihamed}
N.~Arkani-Hamed, L.~J. Hall, H.~Murayama, D.~R. Smith, and N.~Weiner.
\newblock 
\newblock {\em Phys. Rev.}, D64:115011, 2001.

\bibitem{borzumatinomura}
F.~Borzumati and Y.~Nomura.
\newblock 
\newblock {\em Phys. Rev.}, D64:053005, 2001.

\bibitem{casasetal}
J.~A. Casas, J.~R. Espinosa, and I.~Navarro.
\newblock 
\newblock {\em Phys. Rev. Lett.}, 89:161801, 2002.

\bibitem{kitano}
R.~Kitano.
\newblock 
\newblock {\em Phys. Lett.}, B539:102--106, 2002.

\bibitem{arnowittetal}
R.~Arnowitt, B.~Dutta, and B.~Hu.
\newblock 
\newblock {\em Nucl. Phys.}, B682:347--366, 2004.

\bibitem{abelsakis}
S.~A. Abel, A.~Dedes, and K.~Tamvakis.
\newblock 
\newblock {\em Phys. Rev.}, D71:033003, 2005.

\bibitem{admech}
I.~Affleck and M.~Dine.
\newblock {\em Nucl.Phys.B}, 249:361--380, 1985.

\bibitem{balaji}
K.~R.~S. Balaji and R.~H. Brandenberger.
\newblock 
\newblock {\em Phys. Rev. Lett.}, 94:031301, 2005.

\bibitem{drandallt1}
M.~Dine, L.~Randall, and S.~Thomas.
\newblock {\em Nucl. Phys.}, B458:291--326, 1996.

\bibitem{drandallt2}
M.~Dine, L.~Randall, and S.~Thomas.
\newblock {\em Phys. Rev. Lett.}, 75:398--401, 1995.

\bibitem{harveyturner}
J.A Harvey and M.S Turner.
\newblock {\em Phys.Rev.D}, 42(10), 1990.

\bibitem{dreinerross}
H.~K. Dreiner and G.~G. Ross.
\newblock 
\newblock {\em Nucl. Phys.}, B410:188--216, 1993.

\bibitem{tytgat}
N.~Cosme, L.~Lopez~Honorez, and M.~H.~G. Tytgat.
\newblock 
\newblock {\em Phys. Rev.}, D72:043505, 2005.

\bibitem{dmfarrar}
G.~R. Farrar and G.~Zaharijas.
\newblock 
\newblock 2004.
\newblock hep-ph/0406281.

\bibitem{dmbarr1}
S.~M. Barr, R.~S. Chivukula, and E.~Farhi.
\newblock 
\newblock {\em Phys. Lett.}, B241:387--391, 1990.

\bibitem{dmbarr2}
S.~M. Barr.
\newblock 
\newblock {\em Phys. Rev.}, D44:3062--3066, 1991.

\bibitem{dmkaplan}
D.~B. Kaplan.
\newblock 
\newblock {\em Phys. Rev. Lett.}, 68:741--743, 1992.

\bibitem{dmkuzmin}
V.~A. Kuzmin.
\newblock 
\newblock {\em Phys. Part. Nucl.}, 29:257--265, 1998.

\bibitem{dmkitano1}
R.~Kitano and I.~Low.
\newblock 
\newblock {\em Phys. Rev.}, D71:023510, 2005.

\bibitem{dmkitano2}
R.~Kitano and I.~Low.
\newblock 
\newblock 2005.
\newblock hep-ph/0503112.

\bibitem{dmthomas}
S.~D. Thomas.
\newblock 
\newblock {\em Phys. Lett.}, B356:256--263, 1995.

\bibitem{dmkusenko}
A.~Kusenko.
\newblock 
\newblock 1998.
\newblock hep-ph/9901353.

\bibitem{dmhooper}
D.~Hooper, J.~March-Russell, and S.~M. West.
\newblock 
\newblock {\em Phys. Lett.}, B605:228--236, 2005.

\bibitem{cohenkn}
A.~G. Cohen, D.~B. Kaplan, and A.~E. Nelson.
\newblock {\em Ann. Rev. Nucl. Part. Sci.}, 43:27--70, 1993.

\bibitem{troddenrio}
A.~Riotto and M.~Trodden.
\newblock {\em Ann. Rev. Nucl. Part. Sci.}, 49:35--75, 1999.

\bibitem{abazajian}
Kevork Abazajian, George~M. Fuller, and Mitesh Patel.
\newblock 
\newblock {\em Phys. Rev.}, D64:023501, 2001.

\bibitem{allah}
R.~Allahverdi and A.~Mazumdar.
\newblock 
\newblock 2005.
\newblock hep-ph/0512227.

\end{thebibliography}

\end{document}